# Role of the rare-earth doping on the multiferroic properties of BaTiO$_3$: First-principles calculation

A. P. Aslla-Quispe[a,b,*], R. H. Miwa[c], J. D. S. Guerra[a,†]

[a] *Grupo de Ferroelétricos e Materiais Multifuncionais, Instituto de Física, Universidade Federal de Uberlandia, 38408-100, Uberlandia - MG, Brazil*
[b] *Universidad Nacional Tecnológica de Lima Sur, Villa El Savador-Lima, Peru*
[c] *Grupo de Propriedades Eletrônicas e Magnéticas de Moléculas e Sólidos, Instituto de Física, Universidade Federal de Uberlandia, 38408-100, Uberlandia - MG, Brazil*

Ab-initio spin-polarized Density Functional Theory plus U is used to study the electronic and magnetic properties of tetragonal doped barium titanate (Ba$_{1-x}$Eu$_x$O$_3$) system for different europium (Eu$^{3+}$) concentrations. For this study, the Projector Augmented Wave (PAW) method and a Perdew-Zunger (LSDA) approximation, which has been used for the exchange correlation energy, have been considered taking into account a supercell model. In this model, the spin polarization as well as the Hubbard's potential have been used for the correction of the electron-electron Coulomb interactions in the rare-earth ions partially filled *f*-orbitals. The electronic bands-structure reveals that the band-gap energy as well as the dielectric properties decreases with the increase of the doping concentration. On the other hand, the modern theory of polarization also shows that the spontaneous electric polarization increases with the increase of the europium content, whereas the states-density reveals ferromagnetic characteristics (with non-zero total magnetization), without an applied magnetic field, for the Ba$_{1-x}$Eu$_x$O$_3$ system. The magnetic properties also reveal to be strongly dependent on the exchange interaction of the strong localized Eu 4*f*-states in the crystal lattice.

[*] Corresponding author: aaslla@untels.edu.pe (A. P. Aslla-Quispe)
[†] Corresponding author: santos@ufu.br (J. D. S. Guerra)

# 1. Introduction

In recent years, multifunctional materials, which integrate two (or more) fundamental properties, have gained a lot of attention from the scientific community as alternatives to meet the needs of many current and future technological applications [1,2]. In particular, there has been an increasing interest in the study of multiferroic materials, which exhibit simultaneously electrical and magnetic responses [3–6]. It is known that single-phase multiferroics posses intrinsically two (or more) primary ferroic properties, where their corresponding order parameters (electric polarization, magnetization or strain) are switchable by an applied external driven field (electric, magnetic or mechanic) [7,8]. From the fundamental viewpoint, the most commonly studied multiferroic systems are those where the polarization can be affected by a magnetic field, and viceversa (socalled magnetoelectrics), with promissory potential for practical applications, which have motivated both basic and applied researches. However, it is unusual to find materials that are naturally both ferroelectric and magnetic-multiferroics, since in most of the ferroelectric systems, such as the barium titanate ($BaTiO_3$), the ferroelectricity is driven by the hybridization of empty *d*-orbitals of transition metals with occupied *p*-orbitals of the octahedrally coordinated oxygen ions. This mechanism requires empty *d* orbitals and thus cannot lead to a multiferroic behavior. Therefore, there exist very few ferroelectrics that exhibit long-range magnetic order, as well as materials where these two different order parameters coexist and display, in fact, significant coupling.

A ferroelectric material must be an insulator with spontaneous electrical polarization (*P*) as order parameter, generated by structural distortions during the phase transition from a high-symmetry (cubic) phase, at high temperatures, to a lower-symmetry phase (tetragonal, orthorhombic or rhombohedral in case of the $BaTiO_3$ system) at low temperatures. This effect appears as a result of the mismatch in the center of positive and negative charges, thus leading to a permanent electric dipole. In other words, the first requeriment for the ferroelectricity is that the inversion symmetry is broken, while the time-reversal symmetry is preserved [9]. In the tetragonal $BaTiO_3$ (BT), for instance, the ferroelectric property is generated by the small displacement of titanium (Ti) and oxygen (O) atoms along the *z*-axis and the subsequent deformation of the $TiO_6$ oxygen octahedral, which can be observed in the charge distribution map [10,11]. On the other hand, in a ferromagnetic system specific electrical properties are not required, but they show spontaneous magnetization (*M*) as the order parameter caused by a quantum mechanical effect (exchange and super-exchange interactions), which leads the parallel-spins electrons to have lower energy than the electrons with antiparallel-spins below a critical point known as the Curie temperature [12]. In this case, the time-reversal symmetry is broken while the inversion symmetry is preserved [9].

BT system, discovered over the 1940s, is an excellent ferroelectric material with a spontaneous electric polarization around ~26 μC/cm$^2$ [13], below a critical temperature (~403 K). In this case, the alkaline metals (barium) and oxygen ions behaves as completely occupied electronic shells atoms and the strong Ti-O chemical bonding in the crystal promotes the $d^0$ electronic configuration of titanium (3$d$ transition metal), which contradicts the requirement of partially filled $d$-orbitals in transition metals as the main condition for the ferromagnetism [12]. As a result, the magnetic property is suppressed [14]. Nevertheless it is possible to induce ferromagnetic order in the BT system by doping with partially filled $d$ transition metals [15–17]. This phenomenon has been predicted by spin-polarized density functional theory (DFT) calculations [18] and recently observed by experimental measurements in manganese-doped BaTiO$_3$ ceramics [19–21]. In all the cases, the $d$ transition metals substitute generally the titanium (Ti$^{4+}$) ions located at the B-site of the perovskite structure. According to the theory of magnetism, in addition of finding magnetic behavior in transition metals with partially filled $d$-orbitals, it is possible to observe a magnetic behavior in some rare-earth ions with partially filled $f$-orbitals [14]. However, although here have been many published works since the discovery of the barium titanate system, concerning their ferroelectric and multifunctional properties, in which transition metals are used mainly to induce the magnetic properties, only during the last decades with the development of powerful computers, detailed theoretical researches were possible to be carried out. Such studies allow to simulate many interacting particles system by first principles methods calculations. In particular, for obtaining the physical properties of modified crystalline materials with small doping amounts it is necessary to use supercells containing a large number of atoms. In this context, a detailed investigation of the multiferroic properties of rare-earth modified BaTiO$_3$, by considering different Eu$^{3+}$ cation concentrations, is presented in this work. First principles calculations, supported by the Density Functional Theory (DFT) [22,23], have been used to predict the multifunctionality of the studied system. The solution of the Kohn-Sham equations have been carried out by considering an electron subject to an effective potential, which depends on three contributions: the electron-ions interaction, the electron-electron interaction described by means of the Hartree's potential, and the exchange and correlation potential. In our case, in order to better describe the magnetic properties, the theoretical model has been extended to the spin-polarized DFT, where the electronic density has a spin dependence in addition to the spatial position, n$^\sigma$(r), (with σ=↑,↓) [23,24]. On the other hand, due to the inclusion of the Eu$^{3+}$ ion, with strongly correlated $f$- electrons, the standard DFT predicts a metallic behavior. Therefore, in order to correct this behavior, the DFT+U approximation theory has been used, which include the Coulombian repulsion between the strong localized 4*f*-electrons. The spontaneous electric polarization, which characterizes the ferroelectricity, was calculated by considering the quantum geometric-phase, according to the modern theory of the polarization developed by King-Smith and Vanderbilt [25–27]. To the best knowledge of the present authors, no detailed investigation regarding the multifunctional properties of the BT system, which consider spin-polarized Density Functional Theory plus U, has been reported in the literature.

## 2. Methodology and computational procedures

The quantum mechanics first-principles methods are used to study the multiferroics properties of BaTiO$_3$, induced by the inclusion of the Eu$^{3+}$ ions as a dopant element in different concentrations. In order to calculate the electronic and magnetic properties, for most of the cases, the many particle Schrodinger's equation needs to be solved. However, since that method does not provide us a general solution, the spin-polarized Density Functional Theory (DFT) could be used as an alternative to study the structural, electronic and magnetic properties of solids materials and, therefore, the multiferroic properties (including the ferroelectricity and ferromagnetism) can be investigated. In this way, the calculation was performed using the Quantum Espresso software [28,29], where the DFT was implemented with the Ultrasoft Pseudopotential (USP) [30] and the Projector-Augmented Wave (PAW) methods [31]. In order to study the pure ferroelectric perovskites (ABO$_3$) the Local Density Approximation (LDA) for exchange and correlation energy is more efficient than the Generalized Gradient Approximation (GGA) [32]. Therefore, since the GGA Perdew-Burke-Ernzerhof (PBE) [33,34] over estimates the tetragonality (*c/a*) [32,35], we use the spin-polarized Perdew-Zunger (LSDA) approximation [36], including the Hubbard Hamiltonian in the DFT energy functional [37–39]. In this study, the Hubbard potential was considered only for the europium ion, being around 6.90 eV [40,41]. For the calculation of the physical properties, the lattice parameter and atomic positions of the pure BaTiO$_3$ (for tetragonal distortion, with *P4mm* symmetry) were firstly optimized, thus providing the lattice parameter *a*=3.936 Å, tetragonality *c/a*=1.011, and spon*taneous electric polarization P*$_s$=27.368 μC/cm$^2$. The obtained tetragonality and spontaneous electric polarization values are in agreement with the experimental results for *c/a* [42] and *P*$_s$ [13], being in the order of 1.010 and 26.00 μC/cm$^2$, respectively. In order to include the Eu$^{3+}$ ion as substitutional element in the BT structure, different periodic super-lattices have been built, using the optimized atomic positions and lattice parameters for the pure tetragonal

BaTiO$_3$. The barium ion was then substituted by Eu, at different concentrations (*x*), according to the Ba$_{1-x}$Eu$_x$TiO$_3$ formula. The unit-supercells as well as the number of atoms per unit-supercell of the super lattices used in this study are listed in the Table 1, where the 1×1×1 unit-supercell, with a lattice parameter *a*=5.68 Å, represents the unit-cell for Ba$_{0.5}$Eu$_{0.5}$TiO$_3$ composition, as shown in Fig. 1.

**Table 1.** Unit-supercell dimensions, tetragonality c/a and base atoms, considered in this study for different europium concentrations.

| *X* | Supercell | c/*a* | Base atoms |
|---|---|---|---|
| 0.125 | 1×1×4 | 2.85954 | 40 |
| 0.250 | 1×1×2 | 1.42976 | 20 |
| 0.500 | 1×1×1 | 0.71488 | 10 |

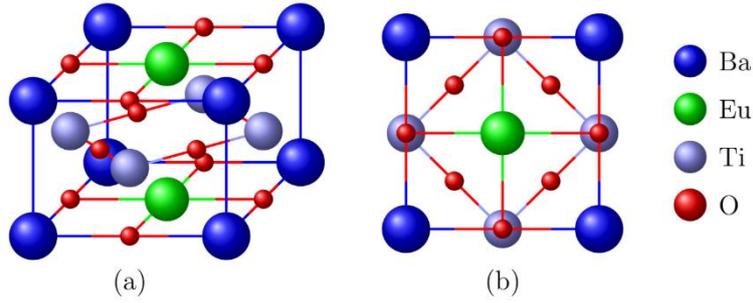

**Figure 1.** Unit-supercell (1×1×1) for the Ba$_{0.5}$Eu$_{0.5}$TiO$_3$ composition.

In order to solve the Kohn-Sham's equations of the spin-polarized DFT, only valence electrons were considered, because the core electrons are strongly bound to the atomic nucleus and do not participate in the chemical bonds. Thus, for each atom in the unit-supercell, the valence bands in our calculations were formed considering 10, 9, 12 and 6 electrons, for the barium ($5s^25p^66s^2$), europium ($4f^65d^16s^2$), titanium ($3s^23p^63d^24s^2$) and oxygen ($2s^22p^4$) ions, respectively. The PAW pseudopotential for Eu$^{3+}$ was generated using the atomic code with [Xe]$4f^65d^16s^2$ as electronic configuration [28] and Troullier-Martins (TM) pseudization procedure [43]. For the calculation of the electronic properties, the atomic positions were optimized after built the unit-supercell, making the total energy to be minimal. For this purpose, the Broyden-Fletcher-Goldfarb-Shanno (BFGS) method was used [44]. After that, to calculate the electronic-density $n^\sigma(r)$, the self-consistent solutions of the Kohn-Sham's equations were performed using 52 Ry kinetic energy cutoff for wavefunctions, 572 Ry kinetic energy for energy density and 6×6×6 Monkhorst-Pack [45] k-points in the first Brillouin zone; finally, the electronic properties were calculated.

## 3. Results and discussion

The BaTiO$_3$ structural unit-cell is formed by the divalent barium and tetravalent titanium cations, which are located at the A- and B- sites, respectively, of the perovskite structure (ABO$_3$), and the divalent oxygen anions in the edge centers. The europium ion has the possibility of being a divalent (Eu$^{2+}$) or trivalent (Eu$^{3+}$) cation. For the trivalent configuration, it has ionic radius around 95 pm and atomic mass of 151.964 g/mol, thus having more chemical affinity to occupy the barium-site rather than the titanium-one. In this context, in this study the barium substitution by europium cation, used as doping element in the BaTiO$_3$, structure has been considered. The amphoteric character presented by the europium ion, where it could also occupy the titanium-site under certain specific conditions, has not been taken into account in this work, and additional analysis regarding this issue, including the influence on the electronic properties, will be investigated in further works.

### 3.1. Ferroelectric properties

The spin-polarized Density Functional Theory was used to optimize the pure BaTiO$_3$ structure (tetragonal symmetry), with the *c*/*a*=1.011 theoretical tetragonality, considering PAW pseudopotential and Perdew-Zunger (LSDA) approximation to exchange correlation energy [36]. The breaking of inversion symmetry, produced by displacement of titanium and oxygen ions, was verified, as can be observed in the total charge distribution shown in the Fig. 2(a). The Ti$^{4+}$ ion is displaced around $\delta z$=-0.0525 Å, while the O$_1$ and O$_2$ oxygen anions are displaced in $\delta z$=0.11354 Å and $\delta z$=0.08679 Å, respectively. The vertical inter-atomic distances Ti-O were found to be around 1.802 Å and 1.907 Å, whose difference is generated by the TiO$_6$ octahedra distortion and produces a non-symmetric charge distribution. As a consequence, the unit-cell has a non-zero electrical dipole moment and the material has a spontaneous electric polarization. The calculations carried out by using the modern theory of polarization [25], implemented in the Quantum Espresso software [28,29], revealed the ferroelectric characteristics of the pure BaTiO$_3$ system, with a spontaneous polarization around 27.368 μC/cm$^2$. The density of states, as well as the spin-up and spin-down bands structures, are shown in the Fig. 3 for the pure and doped BT system. The bands structure, shown in the Fig. 3(a), confirmed the insulating behavior for the pure BaTiO$_3$ system, with a theoretical indirect band-gap energy ($E_g$) of 1.787 eV, which indeed is lower to the reported experimental value for the BaTiO$_3$ system (3.27 eV) [46]; however, it is in agreement with reported theoretical studies for BT using DFT theory [47].

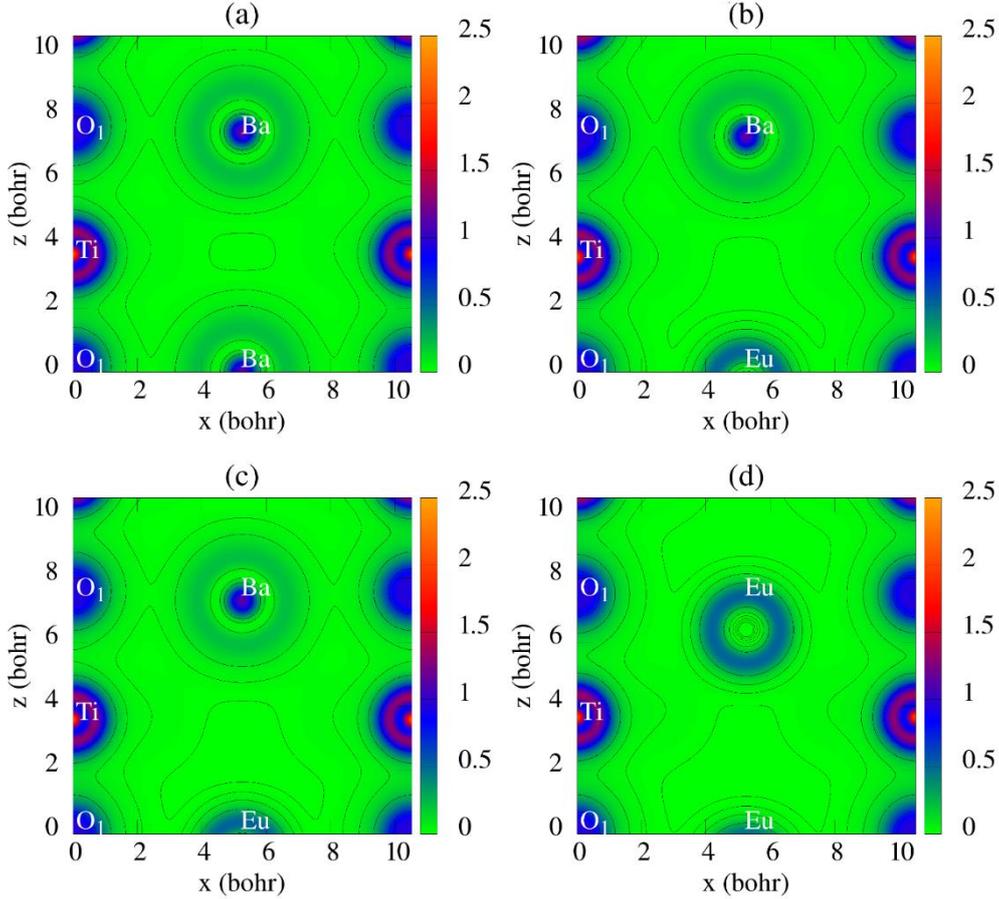

**Figure 2.** Total average charge density configuration in the [010] crystalline plane for: (a) $x$=0.000, (b) $x$=0.125, (c) $x$=0.250 and (d) $x$=0.500.

In order to study the effects of the Eu ions acting as dopants in the BaTiO$_3$ structure, supercells representing the modified BT system (Ba$_{1-x}$Eu$_x$TiO$_3$) with $x$=0.125, 0.250 and 0.500 where built. Then, the system was optimized towards the minimum energy condition and the structural optimizing process revealed the change of the atomic positions with the increase of the Eu concentration. Results, shown in the Table 2, depict that Eu$^{3+}$ moves-down along the $z$-direction, and the Eu cation displacement increases with the increase of $x$. On the other hand, the Ti and O positions also change, varying the deformation of the TiO$_6$ octahedra, as observed in the change of the Ti-O distances involving both the lower and upper oxygens, $\delta$(Ti,O$_{low}$) and $\delta(Ti,O_{up})$, respectively. On the other hand, while $\delta(Ti,O_{low})$ increases, it can be seen that $\delta(Ti,O_{up})$ decreases, with the increase of the Eu content.

**Table 2.** Europium (Eu) and titanium (Ti) displacements, and Ti-O distances along the *z* direction.

| x | $\delta_z$ Eu | $\delta_z$ Ti | $\delta(Ti,O_{low})$ | $\delta(Ti,O_{up})$ |
|---|---|---|---|---|
| 0.000 | – | -0.030599 | 1.875443 | 2.104109 |
| 0.125 | -0.206355 | -0.087762 | 1.914370 | 2.065182 |
| 0.250 | -0.356889 | -0.080009 | 1.915487 | 2.064065 |
| 0.500 | -0.591878 | -0.52900 | 1.942768 | 2.036784 |

A direct consequence of the changes in the atomic positions can be related to the modification in the configuration of the electric charge and the chemical bonds between the ions in the crystalline lattice. It is shown in Fig. 2 the average electric charge distribution over the [010] crystalline plane, with coordinates origin at the (0.0,0.5*a*,0.0) oxygen $O_1$ position, before the optimization process. It is possible to observe, clearly, in Figs. 2(b), 2(c) and 2(d) the change in the position of the $Eu^{3+}$ and oxygen ions, and consequently in the average electric charge distribution.

The inclusion of Eu into the BT structure, however, preserves the breaking of the spatial inversion symmetry and, consequently, affects the electronic and ferroelectric properties, thus promotion the average electric charge reconfiguration with a corresponding change in the electrical dipole moment of the unit-supercell. As a consequence, there is a change in the spontaneous electric polarization of the material in the absence of the external electric field. Table 3 shows the calculated electric and magnetic properties for the studied compositions, including the pure BT. According to the studied configurations, and considering the data reported in Table 3, the composition with a lower Eu concentration (*x*=0.125) showed a spontaneous electric polarization ($P_s$) value lower than that for the pure case, whereas $P_s$ increases for higher Eu concentrations, reaching the maximum value around 89.736 μC/cm² for the *x*=0.50 composition; this later value reveals to be more than twice the obtained value for the *x*=0.25 composition (39.865 μC/cm²). This result can be ascribed to the significant structural distortion and changes in the electronic properties caused by the inclusion of the Eu cation into the BT structure, and also related to the periodic distribution of the Eu in the BaTiO$_3$ crystalline structure, given by $a\hat{\imath} + a\hat{\jmath} + nc\hat{k}$ translation vector, with $n$=3,2,1 for $x$=0.125, 0.250 and 0.500 respectively. This is the main reason for which Fig. 2(d) shows two Eu ions instead of one, as depicted in Figs. 2(b) and 2(c).

**Table 3.** Calculated electrical and magnetic properties for the studied compositions.

| $x$ | Supercell | $E_g$ (eV) | $P_s$ (µC/cm$^2$) | $M$ (µ$_B$/cell) |
|---|---|---|---|---|
| 0.000 | – | 1.787 | 27.368 | 0.00 |
| 0.125 | 1×1×4 | 0.586 | 18.208 | 6.99 |
| 0.250 | 1×1×2 | 1.255 | 39.865 | 7.00 |
| 0.500 | 1×1×1 | 1.315 | 89.736 | 7.00 |

From the technological applications point of view, other important quantity is the band-gap ($E_g$), which is also shown in Table 3. As observed, the effect of the Eu inclusion is to reduce the $E_g$ value, for all the doped compositions, with respect to the pure BT. It is noticed that the lower Eu concentration composition revealed a band-gap energy about 67.21% lower than the pure case, with a semiconductor-like material close behavior, as depicted in Fig. 3(e) (spin-up bands structures for $x$=0.125). However, as shown in Figs. 3(h) and 3(k), the insulating behavior with band-gaps around 29.77% and 26.41% lower than the pure BT, was confirmed for the $x$=0.25 and $x$=0.50 compositions, respectively. It is important to point out that the calculations of the band-gap energies and the spontaneous electric polarization were performed by using the Hubbard potential (U), which describes the Coulomb interaction for the strong localized 4$f$ spin-up and spin-down electrons of the Eu rare-earth ion. This is because the initial DFT calculations, without considering this parameter, produced metallic behaviors for all studied cases, similar to the Mott insulators. According to the obtained results (Table 3), there is a direct correlation between the band-gap energy and the spontaneous electric polarization; that is to say, $P_s$ increases as the Eu concentration increases.

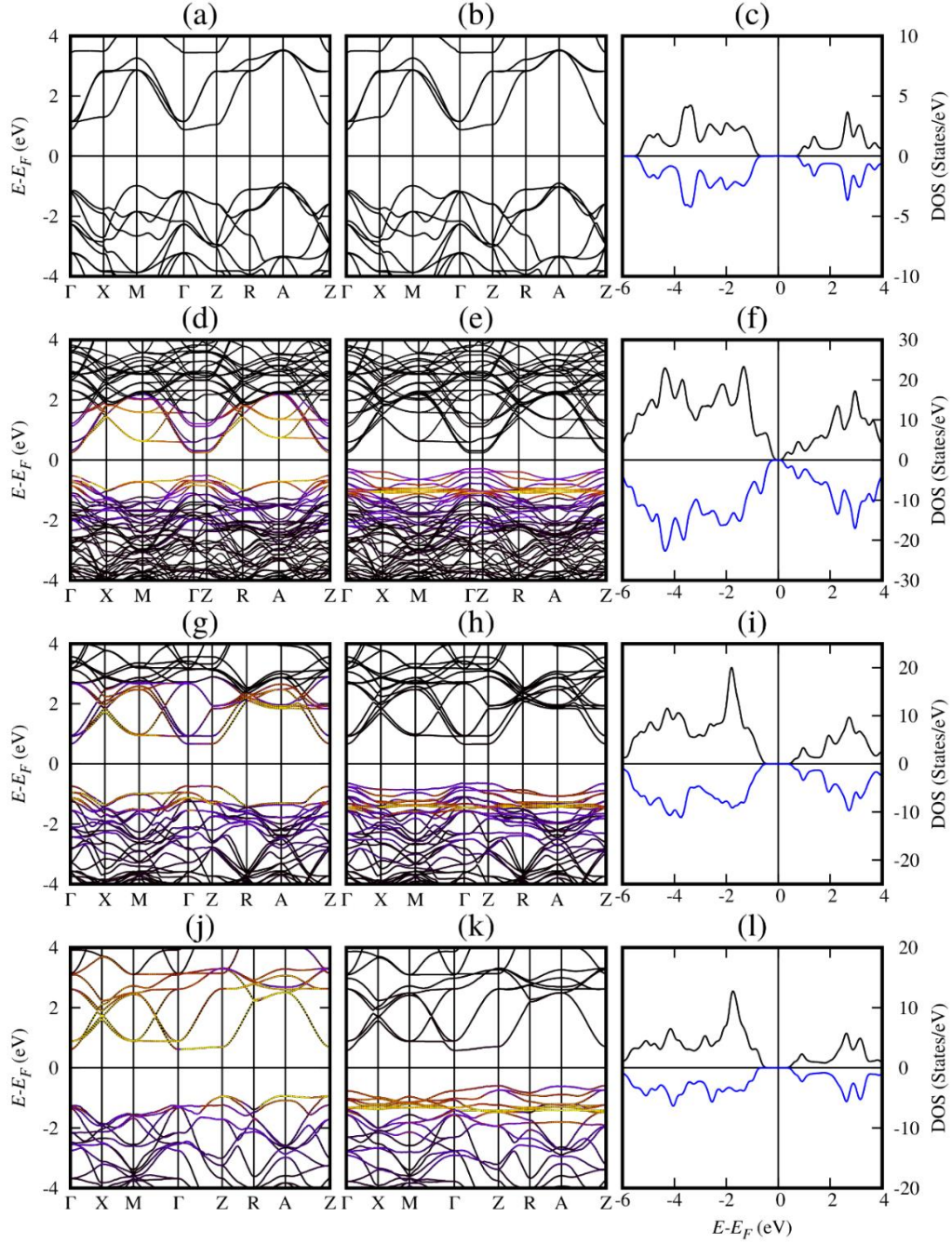

**Figure 3.** Calculated bands structures spin-down (left), spin-up (middle) and total density of states (right), with $x=0.000$ for (a), (b) and (c), $x=0.125$ for (d), (e) and (f), $x=0.250$ for (g), (h) and (i), $x=0.500$ for (j), (k) and (l).

*3.2. Ferromagnetic properties*

From the fundamental point of view, is it known that the ferromagnetic properties of the materials are related to the electronic magnetic moments (spin and orbital components), which that can be analyzed by first principles calculation in multiferroics systems [48–50]. The spin has only a quantum mechanical nature and interacts according to the direct exchange and super-exchange interactions, which take place between electrons due the Pauli's exclusion principle. This condition requires antisymmetric wave function for fermionic indistinguishable particles. The self-consistent calculation, using LSDA plus U, reveals that the total magnetization ($M$) for Eu-doped BaTiO$_3$, calculated by the Eq. (1) [28,51] and listed in Table 3, is around 7 $\mu_B$/cell. The invariance of this result for the three Eu-doped conditions is due to the periodic distribution of the Eu ion in the crystal lattice, maintaining on the $xy$ plane a uniform array, according to the $a\hat{\imath} + a\hat{\jmath}$ translation vector, and varies with the Eu concentration along the $z$-direction, according to $nc\hat{k}$, where $n$=3,2,1 for $x$=0.125, 0.250 and 0.500 respectively.

$$M_T = \int_{cell} \left(\langle n^\uparrow(r)\rangle - \langle n^\downarrow(r)\rangle\right)d^3r \tag{1}$$

Back to the Figs. 3(a) and 3(b), the spin-down and spin-up bands structures, with zero energy in the Fermi level are similar. In Fig. 3(c) the symmetric plot also reveals similar spin-down (blue) and spin-up (black) density of states (DOS). This results confirms the non-magnetic character of the pure BT system. For the doped compositions, however, the LSDA plus U band structures and DOS calculation confirm the ferromagnetic properties of the Ba$_{1-x}$Eu$_x$TiO$_3$ system, with different spin-down and spin-up bands structures, mainly in the valence band, below Fermi level, as shown in Figs. 3(d) and 3(e) for $x$=0.125, in Figs. 3(g) and 3(h) for $x$=0.250, and in Figs. 3(j) and 3(k) for $x$=0.500. Brightest and darkest color scales in these figures represent higher and lowest Eu 4$f$ local states density (LDOS), respectively. On the other hand, results of the total spin-up and spin-down states density also reveals asymmetric representations, as shown in Figs. 3(f), 3(i) and 3(l) for the $x$=0.125, 0.250 and 0.500 compositions, respectively, clearly confirming the non-zero values of the total magnetization given by the Eq. (1). The Coulomb potential (U) certifies that, for the lowest Eu$^{3+}$ concentration ($x$=0.125), the Ba$_{1-x}$Eu$_x$TiO$_3$ system behaves as a ferromagnetic semiconductor with a band-gap around 0.586 eV. For higher europium concentrations, however, the system behaves as a magnetic insulator, with band-gaps around 1.255 eV and 1.315 eV, for the $x$=0.250 and 0.500 concentrations, respectively. It is worth to point out that for all the studied cases the maximum energy of the valence band corresponds to spin-up electron energy levels, while the minimum conduction bands energies correspond to both spin-up and spin-down electrons, as shown the LDOS representations in Figs. 4, 5 and 6.

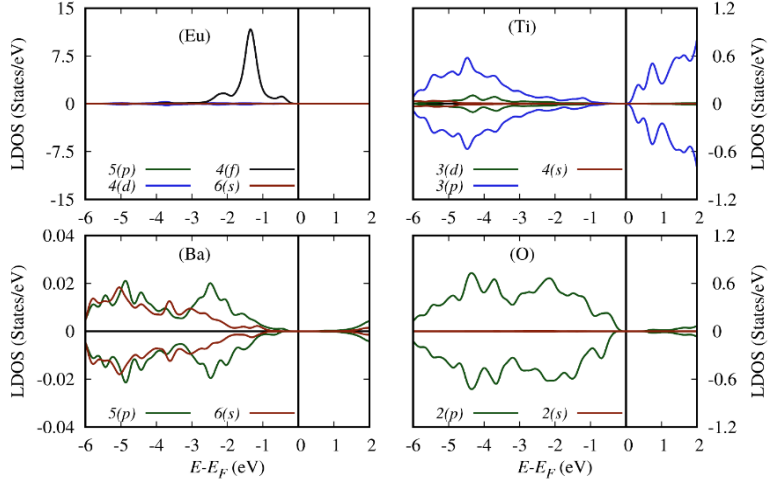

**Figure 4.** Local density of states for $x$=0.125 in arbitrary units (a) europium, (b) titanium, (c) barium and (d) oxygen.

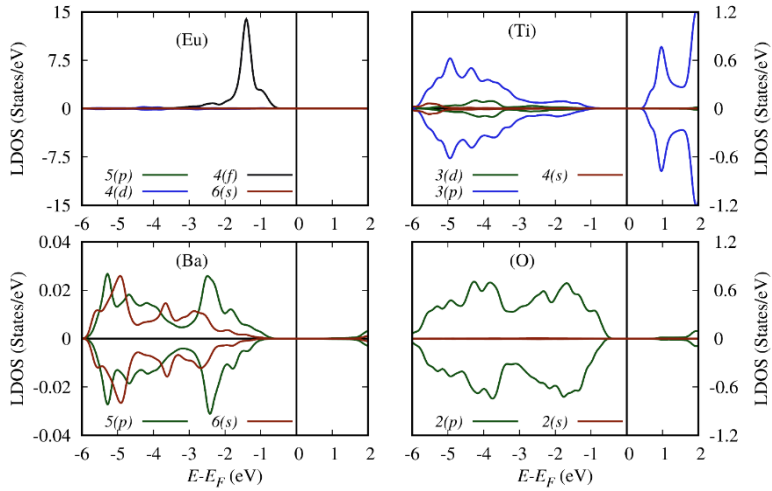

**Figure 5.** Local density of states for $x$=0.25 in arbitrary units (a) europium, (b) titanium, (c) barium and (d) oxygen.

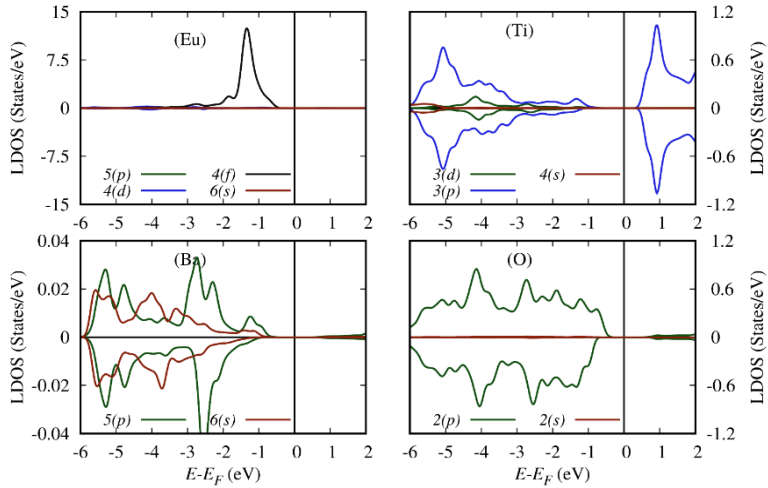

**Figure 6.** Local density of states for $x$=0.500 in arbitrary units (a) europium, (b) titanium, (c) barium and (d) oxygen.

In order to better understand the magnetic contribution for each constituent atom in the $Ba_{1-x}Eu_xTiO_3$ structure, the local densities of states (LDOS) has been used with zero energy in the Fermi level and represented in the Figs. 4, 5 and 6 for the $x$=0.125, 0.250 and 0.500 compositions, respectively. According to the observed results, the cause of the magnetic effects are mainly due to the strong localized $Eu^{3+}$ 4$f$ electrons, which is coherent with the $[Xe]4f^65d^16s^2$ electronic configuration, where the 4f orbitals are partially filled. In the $Ba_{1-x}Eu_xTiO_3$ system, all europium 4$f$ electrons have spin-up, contributing with 6.719 $\mu_B$, 6.6668 $\mu_B$ and 6.5621 $\mu_B$ for the total magnetization, when $x$=0.125, 0.250 and 0.500, respectively. The $Ba^{2+}$, $Ti^{4+}$ and $O^{2-}$ ions present a small contribution for the magnetism of the material, with 0.281 $\mu_B$, 0.3332 $\mu_B$ and 0.4379 $\mu_B$ magnetic moments, for $x$=0.125, 0.250 and 0.500 compositions, respectively. In addition, the LDOS figures revealed that the 4$f$ electrons are localized with $E$-$E_F$ energy values between –2.00 eV and –1.00 eV.

Figures 7, 8 and 9 show the projected density of states (PDOS) over the main orbitals that contributes for the magnetic properties, for the $x$=0.125, 0.250 and 0.500 compositions, respectively, taking into account the LDOS results shown in Figs. 4, 5 and 6. According to the obtained results it is possible to affirm that the effect of the Coulomb potential (U) is to extend the strong localized Eu 4$f$ orbitals, enabling the participation of such electrons in the chemical bonds of the $Ba_{1-x}Eu_xTiO_3$ system. This is because when U=0, the DFT theory predicts a metallic behavior for the $Ba_{1-x}Eu_xTiO_3$ system, with the Eu 4$f$-electrons localized near the Fermi level. On the other hand, the semiconductor behavior observed for the $x$=0.125 composition is caused by electrons in the $f_{y(z^2-x^2)}$, $f_{z(x^2-y^2)}$ and $f_{y^3}$ Eu orbitals, being the $f_{z(x^2-y^2)}$ orbital the more extended one, which contains the most energetic electrons in the valence band. In addition to extend the 4$f$-orbitals, the U potential allows a better understanding of the chemical bonds of the $Eu^{3+}$ cation, with the $Ba^{2+}$, $Ti^{4+}$ and $O^{2-}$ ions, which originate the structural, electronic and magnetic properties of the new doped material.

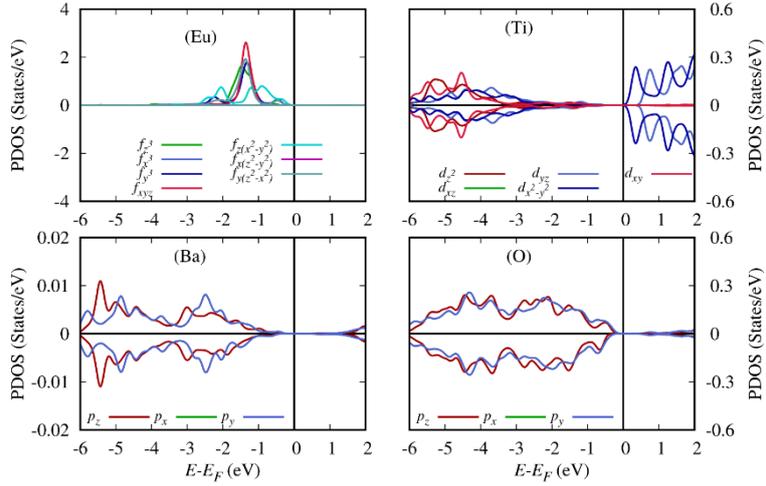

**Figure 7.** Projected density of states for *x*=0.125 in arbitrary units (a) europium 4*f*, (b) titanium 3*d*, (c) barium 5*p* and (d) oxygen 2*p*.

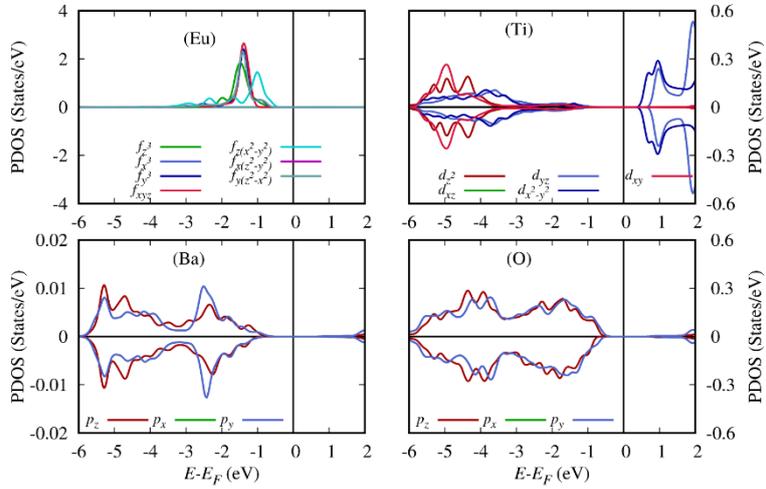

**Figure 8.** Projected density of states for *x*=0.25 in arbitrary units (a) europium 4*f*, (b) titanium 3*d*, (c) barium 5*p* and (d) oxygen 2*p*.

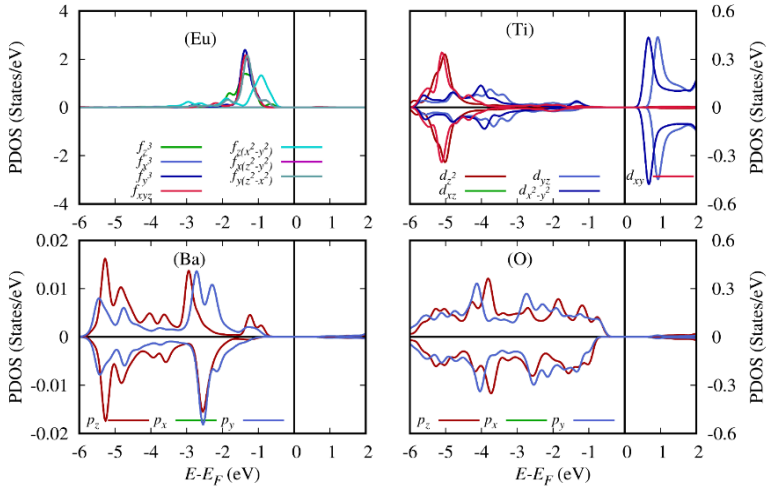

**Figure 9.** Projected density of states for *x*=0.50 in arbitrary units (a) europium 4*f*, (b) titanium 3*d*, (c) barium 5*p* and (d) oxygen 2*p*.

## 4. Conclusions

In summary, the multiferroic properties of $Ba_{1-x}Eu_xTiO_3$, for $x$=0.125, 0.250 and 0.500, were investigated by performing the first-principles LSDA plus U calculation, based on the Perdew-Zunger approximation for the exchange correlation energy and projected augmented wave (PAW) pseudo-potentials. As expected, the pure $BaTiO_3$ system, with tetragonal symmetry, revealed its dielectric behavior with ferroelectric properties. However, the inclusion of the europium rare-earth element into the BT crystalline lattice, promoted the ferromagnetic properties induced by the strong localized $Eu^{3+}$ $4f$-electrons. A ferromagnetic semiconductor behavior was observed for the lowest doping concentration, with a spontaneous electric polarization lower than the pure BT system. For higher rare-earth concentrations, however, an insulator behavior, with spontaneous electric polarization higher than the pure BT system, was obtained. Results revealed the europium and oxygen displacement to be the main factors, rather than the titanium displacement, for the electronic charge configurations and ferroelectric characteristics. The ferromagnetic spontaneous magnetization has shown to be induced mainly by the spin-up europium $4f$ electrons, for all the studied compositions. These results reveal excellent theoretical insights on the multiferroic character of the $BaTiO_3$ system and provide important tools for the developing and understanding of the physical properties of new lead-free multiferroic materials for technological applications.

**Authors' contribution**

A. Aslla-Quispe: Conceptualization, Methodology, Validation, Investigation, Writing - original draft, Visualization. R. H. Miwa: Investigation, Formal analysis, Validation. J. D. S. Guerra: Conceptualization, Writing - review & editing, Visualization, Validation, Funding acquisition, Supervision, Project administration.

**Declaration of Competing Interest**

The authors declare that they have no known competing financial interests or personal relationships that could have appeared to influence the work reported in this paper.

**Acknowledgements**

The authors acknowledge the financial support from CAPES (Finance Code 001), CNPq (303447/2019-2) and FAPEMIG (PPM-00661-16 and APQ-02875-18) Brazilian agencies. Dr. Aslla-Quispe also thanks the CENAPAD-SP for computing calculation facilities.